\documentclass[
 aps,
 ,twocolumn
]{revtex4}
\usepackage{graphicx}
\usepackage{amsmath}
\usepackage{mathrsfs}
\usepackage{color}
\usepackage{tikz}
\usepackage{pgflibraryarrows}
\usepackage{pgflibrarysnakes}
\usetikzlibrary{decorations.pathmorphing}
\usepgflibrary{arrows.meta}

\begin{document}

\title{Spinning up a Time Machine}

\author{T.~C.~Ralph}\email{ralph@physics.uq.edu.au}
\author{Chris Chang}
\affiliation{Centre for Quantum Computation and Communication Technology, \\
School of Mathematics and Physics, University of Queensland, Brisbane, Queensland 4072, Australia}


\date{\today}


\begin{abstract}
{
We present a global metric describing closed timelike curves embedded in Minkowski spacetime. Physically, the metric represents an Alcubierre warp drive on a rotating platform. The physical realizability of such a metric is uncertain due to the exotic matter required to produce it. Never-the-less we suggest that this metric will have applications in more rigorously studying the behavior of quantum fields interacting with closed timelike curves.
}
\end{abstract}

\maketitle

\vspace{10 mm}

\noindent{\it Introduction}:
The worm-hole solution of Morris, Thorne and Yurtsever \cite{morris:1988} generated much interest in closed time like curves (CTCs) because it suggested they may exist in our own universe, as opposed to earlier solutions which had been restricted to different cosmologies \cite{GOD49}. In turn such schemes have been criticised on various grounds \cite{DES92, HAW92}. So far the argument appears undecided \cite{KIM91, VIS03, ORI05, EAR09}. In the end it seems all such discussions require assumptions to be made about the nature of the so far undiscovered complete theory of quantum gravity. Whether CTCs exist or not appears to depend on these assumptions.

Assuming CTCs do exist, a particular interest for some researchers are situations where quantum particles coming from the past in flat-space interact with CTCs before proceeding on to the future in flat space. However, whilst the solution of Ref.\cite{morris:1988} suggests that this sort of interaction is possible it does not explicitly allow one to model such an interaction as there is no explicit connection between the asymptotically flat space time at the two mouths of the worm hole. Whilst one could, in principle, transform between reference frames in order to patch together the desired evolutions, this is challenging in practice. Other solutions are similar, with various exotic spacetimes allowing for CTCs in principle \cite{ac94} but not containing them explicitly, or not allowing an explicit connection between the CTC and chronology preserving objects in flat space, but rather requiring a patching together of different reference frames \cite{ORI05}. Exceptions are recent ad hoc metrics such as in Ref.\cite{FER18}. As a result research into quantum objects interacting with CTCs has used very simple toy models which simply model the CTCs with unusual boundary conditions \cite{DEU91, POL94}.

We will investigate faster-than-light time machines here and be specifically interested in the warp drive metric introduced by Miguel Alcubierre \cite{ac94}. The warp drive  space-time forms a 'bubble' around a region, with contracted space-time in front, and expanded behind. These contractions could push the pocket of space-time at velocities that are not limited by the speed of light. Regular limits on the speed of objects exist due to their motion through space, and so the pocket of space itself
does not experience these same limits. Space-time is flat in the centre of the bubble, hence an observer at this point would be in free-fall. Their geodesic is time-like with no time-dilation with respect to outside of the bubble. As for wormholes, exotic matter/energy is required in order to create the warp drive. Though some work has looked at reducing these requirements \cite{WHI13}, it is not known if the required exotic matter/energy requirements can be met. For the purposes of this analysis we will simply assume they can.

If one is allowed to travel faster than light then time loops are a natural result. Consider the following two journey scenario. In the first journey, with respect to some inertial reference frame, a traveller moves faster than light to a secondary position, some distance away. From the viewpoint of an observer in a different inertial reference frame travelling at some high velocity relative to the first, the temporal order of the space-like separated events describing the departure and arrival of the traveller may be reversed. For this observer, a return journey back from the travellers arrival position to the origin at a finite (but also super-luminal) velocity could allow the traveller to arrive back at their original position at the same time (or earlier) than they had left.

Everett suggested a protocol based on the Alcubierre warp drive and the logic of the previous paragraph that would enable backward time travel \cite{EVE96}. Consider two warp drive bubbles, each defined in inertial reference frames related by a boost and propagating in opposite directions. The traveller uses one bubble to travel superluminally between two points. They then accelerate to match velocities with the reference frame containing the second bubble and use it to travel back to their original position superluminally. In principle if the trajectories of the bubbles and traveller are arranged just right, and the boost is sufficiently large, then the traveller can arrive back at their original position before they left. Intriguing as this is, the arrangement is complicated and requires several changes of reference frame to follow the path of the traveller. It has been suggested that all faster than light time-machines rely on multiple separate journeys in this way \cite{EVE12}. 

Here we propose a single journey time-machine based on faster than light travel. We introduce a single global metric describing backwards in time trajectories for massive particles, allowing for the formation of closed timelike curves and a local metric which describes how events are perceived by local observers. These trajectories are embedded in Minkowski spacetime allowing particles to begin and end their journeys in flat spacetime. To our knowledge metrics of this type have not been described before.

\noindent {\it The Global Metric}:
The coordinates we will use to describe our system are rotating cylindrical coordinates (see Fig.1(a)). The metric for flat space in these coordinates is given by \cite{HAR03},
\begin{eqnarray}\label{rotate}
ds^2 &=& -(1 - \omega^2 r^2) \; dt^2 + 2 \omega r  \; r d \phi dt \; \nonumber \\
&+& r^2 d \phi^2 + dr^2 + d z^2. 
\end{eqnarray}
Here $r$ is the radial coordinate and $\phi$ is the angle in the $x$-$y$ plane. The parameter $\omega$ is the angular velocity around the $z$-axis. An object rotating around the $z$-axis at this angular velocity will be at constant $\phi$ in these coordinates. The time coordinate $t$ can be thought of as that told by a clock at the origin and therefore is not a function of the rotation. When $\omega = 0$ the metric reduces to the usual description of flat space in cylindrical coordinates. The tangential speed of light at a particular radius in terms of these coordinates will be of interest.  The tangential light cones at radius $r$ are given by,
\begin{eqnarray}\label{timemachine}
v_c = 1 \pm \omega r
\end{eqnarray}
The anisotropy of light speed in the tangential directions, in a rotating frame leads to the well-known Sagnac effect in optics.

We now set $\omega = 0$ and consider an Alcubierre warp drive with a circular trajectory in the $x$-$y$ plane which maintains it at a constant radius, $r_s$ (see Fig.1(b)). The corresponding warp drive metric in cylindrical coordinates is given by,
\begin{eqnarray}\label{Alcub}
ds^2 &=& -(1-f^2 v_s^2) \; dt^2 - 2 f v_s \; r d \phi dt \; \nonumber \\
&+& r^2 d \phi^2 + dr^2 + d z^2, 
\end{eqnarray}
where $f \equiv f(d^2_s)$ is a smooth function of the squared distance between the space-time point of interest and a trajectory of constant $r = r_s$ and $z = 0$, and time varying phase, $\phi_s(t)$. Specifically, we can let $d^2_s = r^2 + r_s^2 -2 r r_s cos (\phi - \phi_s(t)) + z^2$ and require that $f(d^2_s)$ is unity for $d^2_s =0$ and smoothly goes to zero as $d^2_s \to R$, where $R$ is some finite positive number. We also define $v_s (t) \equiv r_s {{ d \phi_s}\over{dt}}$. Notice that if we consider incremental translations in the tangential direction and hence set $r d\phi \to dx$ we obtain the original Alcubierre metric \cite{ac94}.
Tangential light cones with $f=1$ are given by,
\begin{eqnarray}\label{timemachine}
v_c = v_s \pm 1.
\end{eqnarray}
We see that a particle moving along with the bubble, i.e. following a circular trajectory with tangential velocity $v_s$, will be following a locally time-like trajectory (i.e. lying between the light cones), even if $v_s > 1$.
\begin{figure}[ht!]
\includegraphics[width=8.0cm]{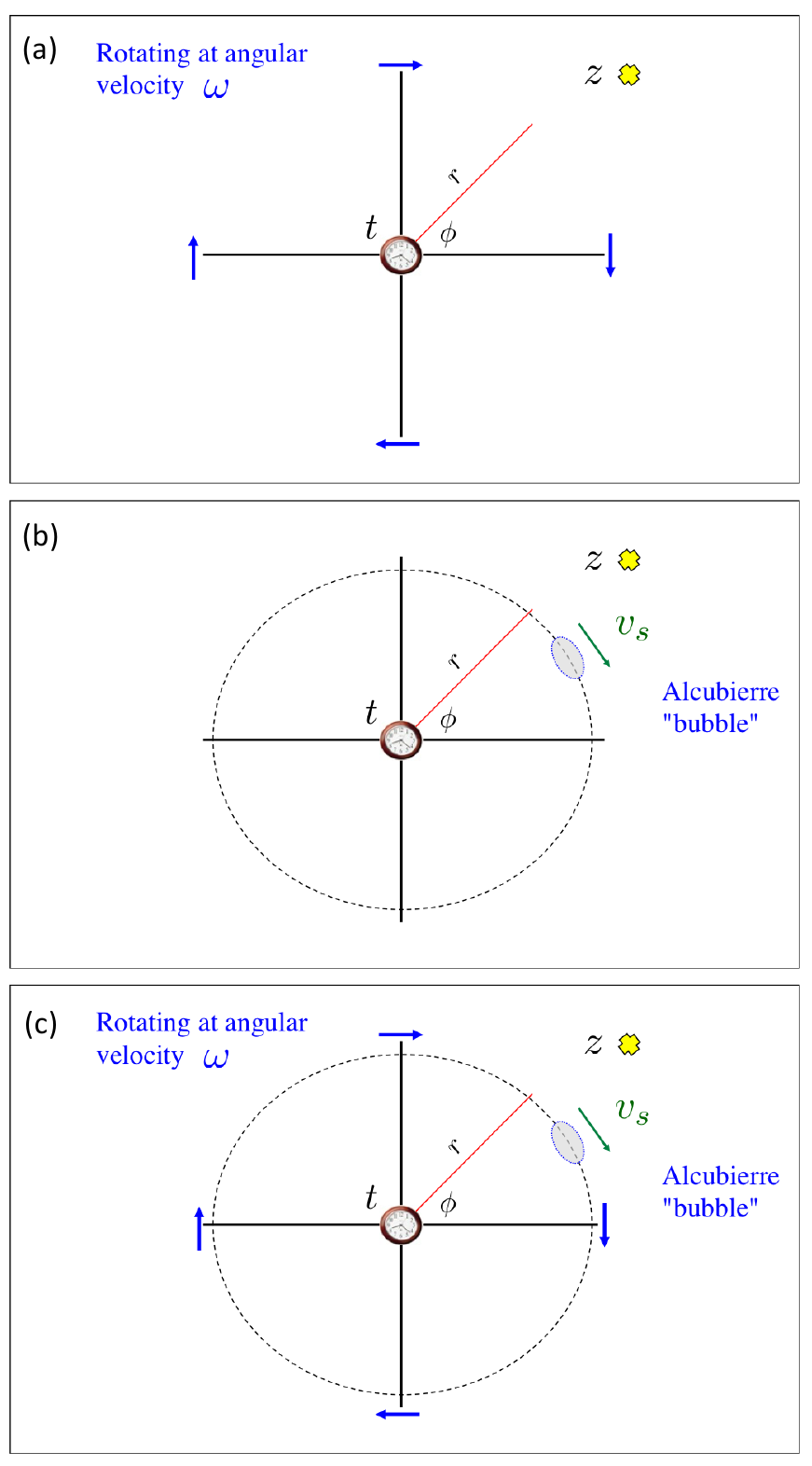}
\caption{\footnotesize Visualisations of the coordinates and ``bubble" for the situations of (a) the rotating metric, (b) the Alcubierre warp drive with a circular trajectory and (c) the Alcubierre warp drive with a circular trajectory on a rotating platform.} 
\label{GeneralCircuit}
\end{figure}
%

We now introduce the time machine metric. We will presently show that this metric represents an Alcubierre warp drive with a circular trajectory as described above, but now on a platform rotating at angular velocity $\omega$ (see Fig.1(c)). A global description of this situation is given by the following metric in our rotating cylindrical coordinates,
\begin{eqnarray}\label{timemachinemetric}
ds^2 &=& -(1-f^2 v_s^2)(1 - \omega^2 r^2) \; dt^2 \nonumber \\ &-& 2(f v_s - \omega r (1-f^2 v_s^2)) \; r d \phi  dt \; \nonumber \\
&+& {{(1 + f v_s \omega r)^2 - \omega^2 r^2}\over{1 - \omega^2 r^2}} r^2 d \phi^2 + dr^2 + d z^2. 
\end{eqnarray}
Tangential light cones with $f=1$ are now given by,
\begin{eqnarray}\label{timemachine}
v_c = {{(v_s \pm 1)(1- \omega^2 r^2)}\over{1+(v_s \pm 1)\omega r}}.
\end{eqnarray}
We will presently show that the condition $(v_s + 1)\omega r < -1$, which leads to a transition of the $+$ solution of $v_c$ from positive to negative, is a necessary condition to allow locally time-like trajectories to exist which propagate backward in time as they travel around the loop.

\noindent{\it The Local Metric}:
We now ask how the metric of Eq.\ref{timemachinemetric} appears locally to an observer on the rotating platform at a radius close to $r_s$. 
First we move from the rotating frame into the lab frame by making the transformation $\phi_z = \phi + \omega t$. The metric becomes
\begin{eqnarray}\label{timemachinemetriclab}
ds^2 &=& -{{(1-(\omega r +f v_s)^2)}\over{(1 - \omega^2 r^2)}} \; dt^2 \nonumber \\ &-& {{2(f v_s + \omega r f^2 v_s^2 + \omega^2 r^2  f v_s)}\over{(1 - \omega^2 r^2)}} \; r d \phi_z  dt \; \nonumber \\
&+& {{(1 + f v_s \omega r)^2 - \omega^2 r^2}\over{1 - \omega^2 r^2}} r^2 d \phi_z^2 + dr^2 + d z^2. 
\end{eqnarray}
Now we move back into the rotating frame, locally at a radius close to $r_s$, by making a tangential Lorentz boost by velocity $-\omega r_s$. The metric reduces to
\begin{eqnarray}\label{timemachinelocal}
ds^2 &=& -(1-f^2 v_s^2) \; dt'^2 - 2 f v_s \; r_s d \phi_z' dt' \; \nonumber \\
&+& r_s^2 d \phi_z'^2 + dr^2 + d z^2, 
\end{eqnarray}
where the dashes indicate boosted coordinates. Notice the similarity between Eq.\ref{timemachinelocal} and Eq.\ref{Alcub} which justifies the claim in the previous section that the metric of Eq.\ref{timemachinemetric} represents an Alcubierre drive on a rotating platform. However, note that the metric of Eq.\ref{timemachinelocal} is only valid close to a constant radius of $r_s$. Also, the range of valid angles $\phi$ we can integrate the metric over should be less than $2 \pi$. If we reach $2 \pi$ then we have come back to where we began and, due the lack of simultaneity in the Lorenz boost, there will be discontinuous jump in the time coordinate. This jump is a well known feature of local metrics in rotating frames \cite{ANA81,RIZ98}, but will take on a special significance here.

In the last section we noted that a particle moving along with the bubble in the metric of Eq.\ref{Alcub}, i.e. following a circular trajectory at radius $r_s$ with tangential velocity $v_s$, will be following a locally time-like trajectory even if $v_s > 1$. This will also be true for the metric of Eq.\ref{timemachinelocal}. It is easily shown that the components of the 4-velocity for such a particle in the local metric are
\begin{eqnarray}\label{timemachinelocalu}
u_L^\alpha = ({{dt'}\over{d\tau}}, {{d\phi_z'}\over{d\tau}}, {{dr}\over{d\tau}}, {{dz}\over{d\tau}}) &=& (1, {{v_s}\over{r}}, 0, 0). 
\end{eqnarray}
Notice, from the time component, we have $dt' = d \tau$, indicating that the clock of the traveller in the bubble remains synchronized with that of an observer (at radius $r_s$) not in the bubble. Making a coordinate transformation to the rotating cylindrical coordinates we find that the components of the particle 4-velocity with respect to the global metric (Eq.\ref{timemachinemetric}) are
\begin{eqnarray}\label{timemachineu}
u_G^\alpha &=& ({{1+v_s \omega r}\over{\sqrt{1- \omega^2 r^2}}}, {{v_s \sqrt{1- \omega^2 r^2}}\over{r}}, 0, 0). 
\end{eqnarray}
The tangential velocity of the particle trajectory in the global metric is
\begin{eqnarray}\label{timemachine}
v_s' = {{r d\phi}\over{dt}} = {{r d\phi}\over{d \tau}}{{d\tau}\over{dt}} =  {{v_s (1- \omega^2 r^2)}\over{1+v_s \omega r}},
\end{eqnarray}
and the relationship between the proper time of a clock carried on this trajectory and co-ordinate time is
\begin{eqnarray}\label{timeG}
d \tau = {{\sqrt{1- \omega^2 r^2}}\over{1+v_s \omega r}} dt.
\end{eqnarray}
In Fig.2 we plot the particle trajectory and the light cones in terms of both the local metric coordinates and the global metric coordinates.
\begin{figure}[ht!]
\includegraphics[width=8.0cm]{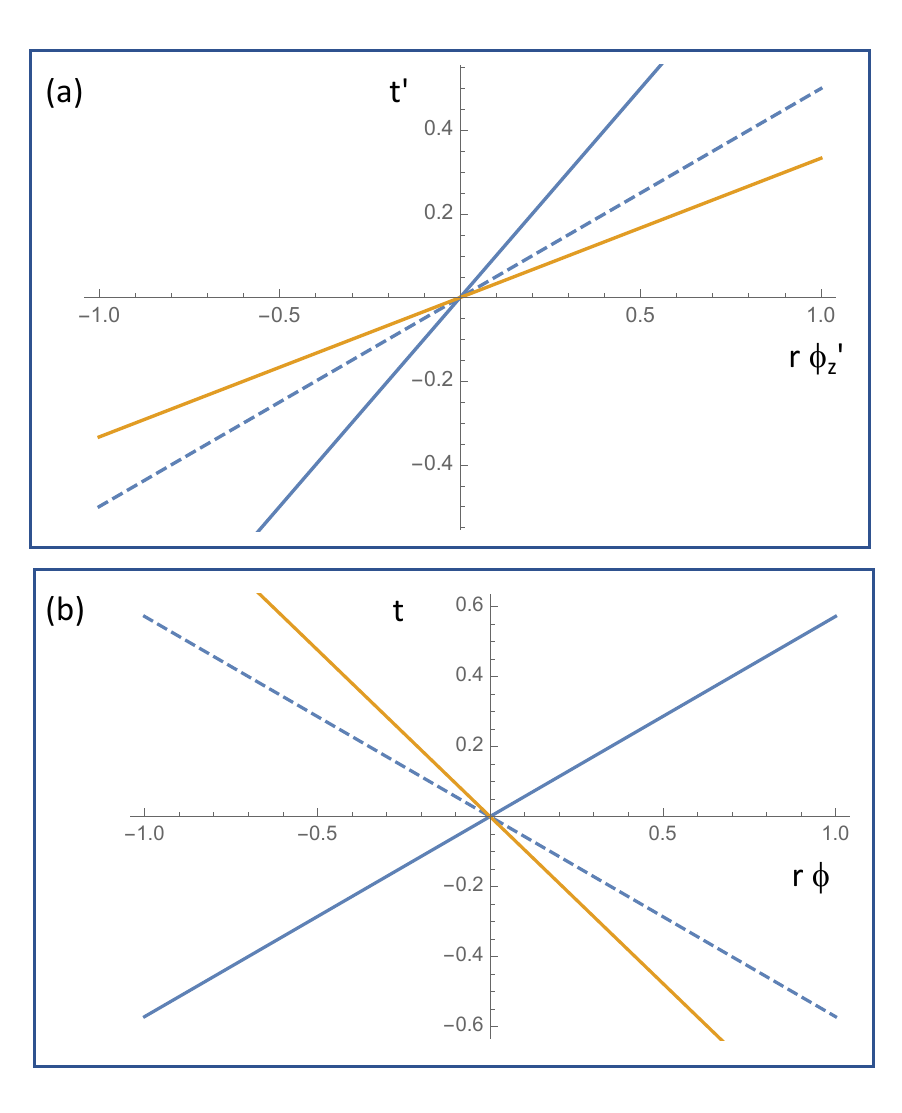}
\caption{\footnotesize Spacetime diagrams showing the tangential lightcones and bubble trajectory in terms of: (a) the local coordinates; and (b) the global coordinates. In the local coordinates the light cones tip over sufficiently to allow the particle trajectory to exceed the speed of light. In the global metric the forward lightcone becomes directed back in time, allowing the particle trajectory to also be directed back in time. The parameters are $v_s = 2$ and $\omega r = -0.75$.} 
\label{spacetimediagram}
\end{figure}

\noindent{\it Time travel examples}:
We now discuss two examples of time travel in this spacetime and analyse them in terms of both the global and local metrics, and use the example parameters in Fig.2 to obtain numerical values for the time taken according to the traveller and the time travelled into the past as observed locally and globally.

Consider a particle moving along with the bubble, i.e. following a trajectory of constant $r = r_s$, $z = 0$ and maintaining a tangential velocity $v_s$ with $\phi(t) = \phi_s(t)$. The 4-velocity of such a particle was discussed in the previous section and is given by Eq.\ref{timemachinelocalu} in the local metric or Eq.\ref{timemachineu} in the global metric. First, let us assume that $v_s(t)$ remains constant over a time interval $t = 0 \to t_f$ which is long enough to allow the particle to make at least one entire circuit at constant velocity. We wish to calculate the proper time interval experienced by the particle and the time interval of an observer who is far enough away to not be affected by the bubble.

First we calculate in the global metric. From Eq.\ref{timemachineu} we have 
\begin{equation}\label{angle}
{{d \phi}\over{d \tau}} = {{v_s \sqrt{1- \omega^2 r_s^2}}\over{r_s}},
\end{equation}
and hence
\begin{eqnarray}\label{tau}
\tau_f &=& \int_0^{2 \pi} {{r_s}\over{v_s \sqrt{1- \omega^2 r_s^2}}} d \phi \nonumber \\
&=& {{2 \pi r_s}\over{v_s \sqrt{1- \omega^2 r_s^2}}}.
\end{eqnarray}
Then from Eq.\ref{timeG} we have
\begin{equation}
t_f = {{2 \pi r_s (1 + v_s \omega r_s)}\over{v_s (1- \omega^2 r_s^2)}}.
\end{equation}
In obtaining these results we are assuming that the initial times at the beginning of the journey are $\tau_i = t_i = 0$. Using the same parameters as in Fig.2 we find that $\tau_f = 4.75 r_s$ while $t_f = -3.59 r_s$. Hence after completing a single circuit the particle has aged by $4.75 {{r_s}\over{c}}$ seconds but arrives back at the origin $3.59 {{r_s}\over{c}}$ seconds in the past.

Now let us perform the same calculation in the local metric. Locally, from Eq.\ref{timemachinelocalu}, we have ${{r_s d \phi}\over{d \tau}} = {{r_s d \phi}\over{d t'}} = v_s$. This implies that $\tau_f = x'_s/v_s$ where $x'_s$ is the locally measured distance around the circuit. Given that the circumference of the circuit is $x_s = 2 \pi r_s$ in the lab frame and taking account of the Lorentz contraction between the lab frame and the local frame, $x_s = \sqrt{1- \omega^2 r_s^2} x'_s$, we obtain the same expression for the proper time as given in Eq.\ref{tau}, as expected.

In order to calculate the time, $t_f'$, told locally by a clock at the origin, we need to account for the fact that due to the Lorentz boosts leading to the local metric there is a lack of simultaneity between the clock at $\phi =0$ and the same clock as described at $\phi = 2 \pi$. This lack of simultaneity, $\Delta t'$ is described by the usual simultaneity relationship, $\Delta t' = x_s v_b/\sqrt{1- v_b^2}$, where $v_b = \omega r_s$ is the boost and $x_s = 2 \pi r_s$ is the separation in the unboosted frame. Given that $d t'/d \tau = 1$ we conclude that the local time will be the proper time plus the lack of simultaneity correction, hence
\begin{equation}
t_f' = \tau + {{2 \pi r_s \omega r_s}\over{\sqrt{1- \omega^2 r_s^2}}}
= {{2 \pi r_s (1 + v_s \omega r_s)}\over{v_s \sqrt{1- \omega^2 r_s^2}}}.
\end{equation}
Accounting for the Lorentz contraction between the global and local frame we see that $t_f' = t_f \sqrt{1- \omega^2 r_s^2}$ as expected.

Now let us consider a more interesting situation in which a particle starts off and finishes in flat space on the rotating platform. An Alcubierre bubble forms at $\phi = t = 0$ and carries the particle around the circuit. The bubble disappears after the circuit, leaving the particle in flat space at the origin again, but at an earlier time. We assume that the velocity of the bubble in the metric changes as a function of proper time, $v_s (\tau) = f(\tau) v_s$, such that $f (0)= 0$ and $f (\tau_a) = 1$, with $f (\tau)$ monotonically increasing between $\tau = 0 \to \tau_a$. We further assume that $f(\tau) =1$ for $\tau_a \leq \tau \leq \tau_f - \tau_a$ and $v_s (\tau) = f (\tau_f -\tau) v_s$ for $\tau_f -\tau_a \leq \tau \leq \tau_f$. Here $\tau_f$ is again the proper time to complete the entire circuit. We again assume that the particle moves so as to remain at the centre of the bubble for the whole trajectory, hence its 4-velocity components in the global coordinates remain those of Eq.\ref{timemachineu} but with $v_s = v_s (\tau)$.

We will work in the global coordinates. Given the symmetry of the trajectory it is sufficient to calculate the time to reach $\phi = \pi$ and then double it. Using Eq.\ref{angle} we can write
\begin{eqnarray}\label{phi}
\phi_1 &=& {{\sqrt{1- \omega^2 r_s^2}}\over{r_s}} \int_0^{\tau_a} f (\tau) v_s  d \tau.
\end{eqnarray}
So after proper time $\tau = \tau_a$ the particle has reached $\phi = \phi_1$. Given that $f(\tau) = 1$ for $\tau_a \leq \tau \leq \pi$ then the proper time for the particle to go between $\phi_1$ and $\pi$ is 
\begin{eqnarray}\label{phipi}
\tau_1 &=& {{r_s}\over{v_s \sqrt{1- \omega^2 r_s^2}}}(\pi -{{\sqrt{1- \omega^2 r_s^2}}\over{r_s}} \int_0^{\tau_a} f (\tau) v_s  d \tau).\nonumber \\
\end{eqnarray}
The total proper time to go around the circuit is then
\begin{eqnarray}\label{phif}
\tau_f &=& 2(\tau_a + \tau_1) = 2 \tau_a - 2 \int_0^{\tau_a} f (\tau) d \tau + {{2 \pi r_s}\over{v_s \sqrt{1- \omega^2 r_s^2}}}.\nonumber \\
\end{eqnarray}
Using Eq.\ref{timeG} it is straightforward to show that the coordinate time for the full circuit is
\begin{eqnarray}\label{phif}
t_f &=&  {{2 \tau_a - 2 \int_0^{\tau_a} f (\tau) d \tau}\over{\sqrt{1- \omega^2 r_s^2}}} + {{2 \pi r_s + v_s \omega r_s}\over{v_s (1- \omega^2 r_s^2)}}.
\end{eqnarray}
It is also straightforward to confirm that a consistent solution emerges from a calculation in the local frame.
To see that time travel is still possible with this protocol let's assume a simple linear ramp function such that $f (\tau) = \tau$ when $0 \leq \tau \leq \tau_a$ with $\tau_a = 1$. Using the parameters from Fig.2 again we find $t_f = (1 - 2 \pi r_s/\sqrt{7})4/\sqrt{7}$. This will be negative, and hence time travel becomes possible, when $r_s > \sqrt{7}/2 \pi \approx 0.42$. If this condition is not met the particle will arrive arrive back at the origin superluminally, but after it left.

If the condition $r_s > \sqrt{7}/2 \pi$ is met the particle ends up in flat space at the origin of the rotating platform {\it before} it left. If it remains there it can become the particle that enters the bubble in the future, thus forming a closed timelike curve. More interestingly, if the particle moves away from the origin it is travelling through its own past and may interact with its past self before following a chronology respecting path away from the bubble and into the future.
%
%
%
%

\indent{\it Discussion and Conclusion}: We can speculate on what a machine able to create the metric of Eq.\ref{timemachinemetric} might look like. Suppose some future technology is able to create a device that via some means can produce the bubble curvature, momentarily, and on command. A large number of such devices could be arranged around the ring and pre-synchronized to fire in sequence in such a way as to create a bubble travelling at varying velocities (including superluminal) around the ring. Imagining this all constructed on a cylindrical platform, then spinning the entire platform at high speeds would result in the desired metric. Obviously this is a gedanken experiment, however it does provides a clear physical picture for attaching to the metric and hence a solid basis upon which to discuss the apparent paradoxes that arise from considering time travel in general relativity.

In conclusion, we have described a global time machine metric which allows one to model particles which enter from a chronology respecting past, interact within a non-chronology respecting interval before exiting into a chronology respecting future. Such situations have been discussed in the literature \cite{FRI90,ECH91}, especially for quantum particles \cite{DEU91, POL94,BAC04,RAL10,LLO11,RIN14} but always with the general relativity abstracted away. The simple form and clear physical interpretation of our metric suggests it can be a key tool for rigorously modelling the dynamics of particles in such situations. Such abstracted versions also arise in research into indefinite causal structures \cite{OGN12,CHI13} and could also benefit from the metric discussed here. Ultimately we hope this will lead to a better understanding of the consequences for the union of quantum mechanics and general relativity of the potential existence of such spacetimes.

\indent{\it Acknowledgements}: We thank Alex Tremayne for useful discussions. This work was partially supported by the by the Australian Research
Council Centre of Excellence for Quantum Computation and Communication Technology (Project No.CE110001027).

\end{document}